\documentclass[aps,preprint,showpacs,groupedaddress,floatfix]{revtex4}
\usepackage{graphicx}
\usepackage{dcolumn}
\usepackage{bm}

\newcommand{\lp}{\left}
\newcommand{\rp}{\right}

\newcommand{\ket}[1]{\left| #1 \right\rangle}

\begin{document}

\title{$\beta$-decay rates of r-process nuclei in the relativistic quasiparticle
random phase approximation}
\author{T. Nik\v si\' c}
\author{T. Marketin}
\author{D. Vretenar}
\affiliation{Physics Department, Faculty of Science, University of Zagreb, 
Croatia}
\author{N. Paar}
\affiliation{
Institut f\" ur Kernphysik, Technische Universit\" at Darmstadt, 
Schlossgartenstrasse 9,
D-64289 Darmstadt, Germany}
\author{P. Ring}
\affiliation{Physik-Department der Technischen Universit\"at M\"unchen, 
D-85748 Garching,
Germany}
\date{\today}

\begin{abstract}
The fully consistent relativistic proton-neutron quasiparticle
random phase approximation (PN-RQRPA) 
is employed in the calculation of 
$\beta$-decay half-lives of neutron-rich nuclei in the 
N$\approx$50 and N$\approx$82 regions. A new density-dependent
effective interaction, with an enhanced value of the nucleon
effective mass, is used in relativistic Hartree-Bogoliubov 
calculation of nuclear ground states and in the particle-hole
channel of the PN-RQRPA. The finite range Gogny D1S interaction
is employed in the $T=1$ pairing channel, and the model 
also includes a proton-neutron particle-particle interaction.
The theoretical half-lives reproduce the experimental data 
for the Fe, Zn, Cd, and Te isotopic chains, but overestimate the
lifetimes of Ni isotopes and predict a stable $^{132}$Sn.   
\end{abstract}

\maketitle

\bigskip \bigskip
\section{\label{secI}Introduction}
The latest theoretical and 
computational advances in nuclear structure modeling have also 
had a strong impact on nuclear astrophysics. More and more often 
calculations of stellar nucleosynthesis, nuclear aspects of 
supernova collapse and explosion, and neutrino-induced reactions, 
are based on microscopic global predictions for the nuclear 
ingredients, rather than on oversimplified 
phenomenological approaches. The nuclear input for astrophysics 
calculations necessitates the properties of 
thousands of nuclei far from stability, including the 
characteristics of strong, electromagnetic and weak interaction 
processes. Most of these nuclei, especially on the neutron-rich 
side, are not accessible in experiments and, therefore, many 
nuclear astrophysics calculations crucially depend on accurate 
theoretical predictions for the nuclear masses, bulk properties, 
nuclear excitations, ($n,\gamma$) and ($\gamma,n$) rates, 
$\alpha$- and $\beta$-decay half-lives, fission probabilities, 
electron and neutrino capture rates, etc. 

The path of the $r$-process nucleosynthesis runs through regions 
of very neutron-rich nuclei. $\beta$-decays are particularly 
important because they generate 
elements with higher Z-values, and set the time scale of
the $r$-process. Except for a few key nuclei, however, 
$\beta$-decays of $r$-process nuclei have to be determined 
from nuclear models. One of the crucial questions for structure
models is, therefore, a consistent microscopic calculation 
of $\beta$-decay far from stability. Two microscopic 
approaches have been successfully applied in large-scale modeling 
of weak interaction rates of neutron-rich nuclei: 
the shell model and the proton-neutron quasiparticle random 
phase approximation (PN-QRPA). Shell model applications in
nuclear astrophysics have recently been reviewed in 
Ref.~\cite{LM.03}. In the calculation of 
$\beta$-half lives, in particular,
the principal advantage of the shell model is the ability 
to take into account the detailed structure of the $\beta$-strength
functions. In addition to large-scale shell model predictions 
for the half-lives of waiting-point nuclei at $N=50, 80, 126$ \cite{LM.03},
the no-core shell model \cite{CNOV.02} and the shell model 
embedded in the continuum \cite{MONP.02} have recently been 
applied to $\beta$-decay of light nuclei. Large configuration 
spaces, however, prevent systematic applications of the shell model to 
heavy nuclei along the r-process path. 

When compared to the shell model, important advantages of an 
QRPA approach based on the microscopic self-consistent mean-field
framework, include the use of global effective nuclear interactions 
and the treatment of arbitrarily heavy systems. Fully self-consistent
QRPA calculations of weak interaction rates of neutron-rich nuclei, 
however, have only recently been reported. In Ref.~\cite{Eng.99}, 
in particular, a fully self-consistent QRPA has been formulated in 
Hartree-Fock-Bogoliubov (HFB) canonical single-particle basis. 
The canonical basis diagonalizes the 
density matrix and 
describes both the bound states and the positive-energy single-particle
continuum. The formulation of the RQRPA in the canonical basis is
particularly convenient because, in order to describe transitions to
low-lying states in weakly bound nuclei, the
two-quasiparticle configuration space must include states with both nucleons
in the discrete bound levels, states with one nucleon in a bound level and
one nucleon in the continuum, and also states with both nucleons in the
continuum. The HFB+QRPA model based on Skyrme interactions, and with 
the inclusion of a finite-range residual proton-neutron particle-particle
interaction, has been applied in the calculation of $\beta$-decay 
rates for spherical neutron-rich r-process waiting-point nuclei~\cite{Eng.99}.
Based on a different energy-density functional, 
the self-consistent HFB plus continuum QRPA framework has been 
employed in a systematic calculation of 
the allowed and first-forbidden $\beta$-decay rates for the $r$-process 
nuclei near $N=50, 80, 126$ \cite{Bor.03}. It has been shown 
that the effect of the high-energy first-forbidden transitions 
is crucial for $Z\geq 50$, $N\approx 82$, and especially 
in the $N=126$ region.

In this work we report the first application of the relativistic 
PN-QRPA in the calculation of $\beta^-$-decay rates for 
neutron-rich nuclei. Self-consistent relativistic mean-field models
have been very successfully applied to
nuclear structure, not only in nuclei along the valley of $\beta$-stability, 
but also in exotic nuclei with extreme isospin 
values and close to the particle drip lines.
The relativistic mean-field framework has recently been extended to 
include medium dependent meson-nucleon vertices. 
One way to determine the functional form of the meson-nucleon 
vertices is from in-medium Dirac-Brueckner interactions, obtained
from realistic free-space NN interactions. This microscopic method
represents an {\it ab initio} description of nuclear matter and
finite nuclei. Another approach is purely phenomenological, 
with the density dependence  
for the $\sigma$, $\omega$ and $\rho$ meson-nucleon
couplings adjusted to properties of nuclear matter and 
a set of spherical nuclei. The resulting density-dependent 
meson-nucleon couplings are in qualitative agreement with those
obtained in the microscopic approach. 
In Ref. \cite{NVFR.02} we have extended the relativistic
Hartree-Bogoliubov (RHB) model to include medium-dependent
vertex functions. A phenomenological effective interaction,
denoted DD-ME1, was adjusted to properties
of nuclear matter and finite nuclei, and tested in the analysis 
of the equations of state for symmetric
and asymmetric nuclear matter, of ground-state properties of the
Sn and Pb isotopic chains \cite{NVFR.02}, and deformed nuclei \cite{NVLR.04}.   
It has been shown that, in comparison with standard
non-linear meson self-interactions, relativistic Lagrangians with an
explicit density dependence of the meson-nucleon couplings provide
an improved description of asymmetric nuclear matter, neutron
matter and nuclei far from stability. The relativistic
random-phase approximation (RRPA), based on effective Lagrangians
characterized by density-dependent meson-nucleon vertex functions,
has been derived in Ref. \cite{NVR.02}. In Ref.~\cite{Paa.03}
the relativistic QRPA has been formulated in the canonical
single-nucleon basis of the RHB model, and the corresponding 
proton-neutron relativistic QRPA (PN-RQRPA) has recently been
employed in studies of isobaric analog resonances and Gamow-Teller 
resonances in spherical nuclei \cite{VPNR.03,Paa.04}. 

Our initial attempt to calculate $\beta^-$ decay rates of r-process 
nuclei by simply employing the DD-ME1 interaction in the RHB plus
PN-RQRPA framework was not successful. In general, the resulting 
half-lives were more than an order of magnitude longer than the 
empirical values. The reason is that DD-ME1, as well as most other 
successful relativistic mean-field interactions, has a 
relatively low effective nucleon mass, especially when 
compared to effective masses of Skyrme
forces. This means that, because the density of states around 
the Fermi surface is low, 
in a self-consistent relativistic 
QRPA calculation of $\beta$-decay the transition energies 
are low, and this result in long lifetimes. 
In some cases one could improve 
the results by adjusting the strength of the $T=0$ pairing 
interaction, which is not really known, but this is not the 
solution of the problem, and in any case it cannot be applied
to doubly closed-shell nuclei. The problem of adjusting the 
effective nucleon mass in relativistic mean-field models is 
somewhat complicated, because this quantity is also related to 
the energy spacings between spin-orbit partner states. Since 
the latter are empirically well determined, the effective 
nucleon mass cannot be increased without extending the 
standard relativistic mean-field framework. In order to be 
able to calculate $\beta$-decay half-lives, in this work 
we adjust a new relativistic mean-field density-dependent 
interaction with a higher value of the 
effective nucleon mass, by including an isoscalar tensor-coupling
term in the model Lagrangian. In Sec. \ref{secII} we outline
the theoretical framework and derive the new effective 
interaction DD-ME1*. In Sec. \ref{secIII} the 
relativistic PN-QRPA is employed in the calculation of 
$\beta$-decay half-lives in the regions N$\approx$50 and 
N$\approx$82. Sec. \ref{secIV} contains the summary, 
concluding remarks and an outlook of future applications.

\section{\label{secII}Outline of the theoretical framework}
\subsection{Proton-neutron relativistic quasiparticle random-phase approximation}

The relativistic QRPA has been derived in 
Ref.~\cite{Paa.03}, in the canonical
single-nucleon basis of the relativistic Hartree-Bogoliubov (RHB) model. The
RHB model presents the relativistic extension of the Hartree-Fock-Bogoliubov
framework, and it provides a unified description of $ph$ and $pp$
correlations. In
the RHB framework the ground state of a nucleus can be written either in the
quasiparticle basis as a product of independent quasi-particle states, or in
the canonical basis as a highly correlated BCS-state. By definition, the
canonical basis diagonalizes the density matrix and it is always localized.
It describes both the bound states and the positive-energy single-particle
continuum. The matrix equations of the QRPA take a particularly 
simple form in the canonical basis, including only the matrix
elements $V_{\kappa ^{{}}\lambda ^{\prime }\kappa ^{\prime }\lambda
^{{}}}^{ph}$ of the residual 
$ph$-interaction, and the matrix elements $V_{\kappa
^{{}}\kappa ^{\prime }\lambda ^{{}}\lambda ^{\prime }}^{pp}$ 
of the pairing $pp$-interaction, as well as certain combinations of the 
occupation factors $u_{\kappa }$, $v_{\kappa }$ of the canonical 
single-nucleon states.
The RQRPA configuration space includes the Dirac sea of negative energy
states. In addition to the configurations built from two-quasiparticle
states of positive energy, the RQRPA configuration space must also contain
pair-configurations formed from the fully or partially occupied states of
positive energy and the empty negative-energy states from the Dirac sea. The
inclusion of configurations built from occupied positive-energy states and
empty negative-energy states is essential for current conservation and the
decoupling of spurious states. In recent applications of the
relativistic (Q)RPA it has been shown that the fully consistent inclusion of
the Dirac sea of negative energy states in the R(Q)RPA configuration space is
essential for a quantitative comparison with the experimental excitation
energies of giant resonances~\cite{Paa.03,Vre.00,Ring.01}.

The RHB+RQRPA model is fully self-consistent.
For the interaction in the particle-hole channel effective Lagrangians with
nonlinear meson self-interactions 
or density-dependent meson-nucleon couplings 
are used, and pairing correlations are
described by the pairing part of the finite range Gogny interaction. Both in
the $ph$ and $pp$ channels, the same interactions are used in the RHB
equations that determine the canonical quasiparticle basis, and in the
matrix equations of the RQRPA. This is very important, because the energy
weighted sum rules are only satisfied if the pairing interaction is
consistently included both in the static RHB and in the dynamical RQRPA
calculations. In both channels the same strength parameters of the 
interactions are used in the RHB and RQRPA calculations.

In Ref.~\cite{Paa.04} we have derived the matrix equations of 
the corresponding proton-neutron relativistic QRPA (PN-RQRPA) 
for an effective Lagrangian characterized by 
density-dependent meson-nucleon couplings.
The PN-RQRPA has been applied in a study of charge-exchange modes:
isobaric analog resonances and Gamow-Teller resonances.
The model includes both the $T=1$ and $T=0$ pairing channels, 
and presents a relativistic extension of the fully-consistent 
proton-neutron QRPA that has been formulated in Ref.~\cite{Eng.99},
and employed in an analysis of $\beta$-decay rates of $r$-process nuclei.
In order to build a complete basis, the model space also 
includes configurations built from occupied positive-energy states and
empty negative-energy states. It has been shown recently \cite{Ma.03,Paa.04},
that the total GT strength in the nucleon sector is reduced by 
$\approx 12$\% in nuclear matter, and by $\approx 6$\% in finite 
nuclei when compared to the Ikeda sum rule \cite{Ike.63}
\begin{equation}
\left(  S_{\beta^{-}}^{GT}-S_{\beta^{+}}^{GT}\right)  =3(N-Z),\label{gtsrule}%
\end{equation}
where $S_{\beta^{\pm}}^{GT}$ denotes the total sum of Gamow-Teller
strength for the $\beta^{\pm}$ transition. The reduction has been attributed
to the effect of Dirac sea negative-energy states, i.e. the missing 
part of the sum rule is taken by configurations formed from occupied 
states in the Fermi sea and empty negative-energy states in the Dirac sea.

We consider transitions between the $0^+$ ground state of a 
spherical even-even
parent nucleus and the $1^+$ excited state of the 
corresponding odd-odd daughter
nucleus, which are induced by the Gamow-Teller operator $\bm{\sigma}\tau_-$.
The matrix equations of the PN-RQRPA read

\begin{equation} \left( \begin{array} [c]{cc}
A & B\\
B^{^{\ast}} & A^{^{\ast}}
\end{array}
\right)  \left( \begin{array} [c]{c}
X^{\lambda }\\
Y^{\lambda }
\end{array}
\right)  =E_{\lambda}\left( \begin{array} [c]{cc}
1 & 0\\
0 & -1
\end{array}
\right)  \left( \begin{array} [c]{c}
X^{\lambda }\\
Y^{\lambda }
\end{array}\right) \; . 
\label{pnrqrpaeq}
\end{equation}
The matrices $A$ and $B$ are defined in the canonical basis \cite{Rin.80}

\begin{eqnarray}
A_{pn,p^\prime n^\prime} &=& H^{11}_{pp^\prime}\delta_{nn^\prime} +
  H^{11}_{nn^\prime}\delta_{pp^\prime}  \nonumber \\ & & +
\lp( u_p v_n u_{p^\prime} v_{n^\prime} + v_p u_n v_{p^\prime} u_{n^\prime}\rp)
 V_{pn^\prime n p^\prime}^{ph } + 
\lp( u_p u_n u_{p^\prime} u_{n^\prime} + v_p v_n v_{p^\prime} v_{n^\prime}\rp) 
 V_{pn p^\prime n^\prime}^{pp } \nonumber \\
B_{pn,p^\prime n^\prime} &=& 
\lp( u_p v_n v_{p^\prime} u_{n^\prime} + v_p u_n u_{p^\prime} v_{n^\prime}\rp)
 V_{pp^\prime n n^\prime}^{ph } \nonumber \\ & &- 
\lp( u_p u_n v_{p^\prime} v_{n^\prime} + v_p v_n u_{p^\prime} u_{n^\prime}\rp) 
 V_{pn p^\prime n^\prime}^{pp } \; .
\label{abmat}
\end{eqnarray}
Here $p$, $p^\prime$, and $n$, $n^\prime$ denote proton and 
neutron quasiparticle
canonical states, respectively, 
$V^{ph}$ is the proton-neutron particle-hole residual 
interaction in the $1^+$ channel,
and $V^{pp}$ is the corresponding particle-particle interaction.
The canonical basis diagonalizes the density matrix, and 
the occupation amplitudes $v_{p,n}$ are the corresponding eigenvalues. 
However, the canonical basis does not diagonalize the Dirac single-nucleon
mean-field Hamiltonian $\hat{h}_{D}$ or the pairing field $\hat{\Delta}$,
and therefore the off-diagonal matrix elements $H^{11}_{nn^\prime}$ and
$H^{11}_{pp^\prime}$ appear in Eq. (\ref{abmat}):
\begin{equation}
H_{\kappa \kappa^\prime}^{11}=(u_{\kappa }u_{\kappa^\prime }
-v_{\kappa }v_{\kappa^\prime
})h_{\kappa \kappa^\prime }-(u_{\kappa }v_{\kappa^\prime }+
v_{\kappa }u_{\kappa^\prime
})\Delta _{\kappa \kappa^\prime }\;.
\label{H11}
\end{equation}
For each energy $E_{\lambda}$, $X^{\lambda}$ and $Y^{\lambda}$
in Eq. (\ref{abmat})
denote the corresponding forward- and backward-going QRPA amplitudes,
respectively. The total strength for the transition 
between the ground state of the even-even
(N,Z) nucleus and a $1^+$ state of the odd-odd (N-1,Z+1) 
nucleus, induced by the Gamow-Teller operator, reads
\begin{equation}
B_{\lambda} = \lp| \sum_{pn} <p||\bm{\sigma}\tau_-||n> 
\lp( X_{pn}^{\lambda } u_p v_n - Y_{pn}^{\lambda }v_p u_n \rp) \rp|^2\; .
\label{strength-}
\end{equation}

The spin-isospin-dependent interaction terms are 
generated by the $\pi$- and $\rho$-meson exchange.
Although the direct one-pion contribution to the nuclear ground state 
vanishes at the mean-field level because of parity conservation, 
it has to be included in the calculation of spin-isospin excitations.
The particle-hole residual interaction of the PN-RQRPA is derived from the 
Lagrangian density 
\begin{equation}
\mathcal{L}_{\pi + \rho}^{int} = 
      - g_\rho \bar{\psi}\gamma^{\mu}\vec{\rho}_\mu \vec{\tau} \psi 
      - \frac{f_\pi}{m_\pi}\bar{\psi}\gamma_5\gamma^{\mu}\partial_{\mu}
        \vec{\pi}\vec{\tau} \psi \; . 
\label{lagrres}	
\end{equation}
Vectors in isospin space are denoted by arrows, and boldface symbols 
will indicate vectors in ordinary three-dimensional space.

The coupling between the $\rho$-meson and the nucleon is assumed to be 
a vertex function of the baryon density. In Ref.~\cite{NVR.02} it has been
shown that the explicit density dependence of the meson-nucleon couplings
introduces additional rearrangement terms in the residual two-body 
interaction of the RRPA, and that their 
contribution is essential for a quantitative description of excited
states. However, since the rearrangement terms include the corresponding 
isoscalar ground-state densities, it is easy to see that they are absent in the 
charge exchange channel, and the residual two-body interaction reads
\begin{eqnarray}
V(\bm{r}_1,\bm{r}_2) &=& \vec{\tau}_1\vec{\tau}_2 (\beta \gamma^\mu)_1
      (\beta \gamma_\mu)_2 g_\rho[\rho_v(\bm{r}_1)] g_\rho[\rho_v(\bm{r}_2)]
      D_\rho (\bm{r}_1,\bm{r}_2) \nonumber \\
      &-&\left(\frac{f_\pi}{m_\pi}\right)^2\vec{\tau}_1\vec{\tau}_2
      (\bm{\Sigma}_1\bm{\nabla}_1)(\bm{\Sigma}_2\bm{\nabla}_2)
      D_\pi (\bm{r}_1,\bm{r}_2)\;.
\end{eqnarray}
$D_{\rho (\pi)}$ denotes the meson propagator
\begin{equation}
 D_{\rho (\pi)} = \frac{1}{4\pi}
        \frac{e^{-m_{\rho (\pi)}|\bm{r}_1-\bm{r}_2|}}{|\bm{r}_1-\bm{r}_2|}\;, 
\end{equation}	   

and 
\begin{equation}
\bm{\Sigma} = \left(
\begin{array}
[c]{cc}%
\bm{\sigma} & 0\\
0 & \bm{\sigma}%
\end{array}
\right)  \;. \label{bigsigma}%
\end{equation}
For the $\rho$-meson coupling we adopt the functional form used in
the DD-ME1 density-dependent effective interaction~\cite{NVFR.02}
\begin{equation}
g_\rho (\rho_v) = g_\rho(\rho_{sat})exp[-a_\rho (x-1)]\;,
\end{equation}
where $x=\rho_v / \rho_{sat}$, and $\rho_{sat}$ denotes the saturation vector 
nucleon density in symmetric nuclear matter.
For the pseudovector pion-nucleon coupling we use the standard values 
\begin{equation}
m_{\pi}=138.0~{\rm MeV}~~~~\;\;\;\;\frac{\;f_{\pi}^{2}}{4\pi}=0.08\;.
\end{equation}
The derivative type of the pion-nucleon coupling necessitates the  
inclusion of the zero-range
Landau-Migdal term, which accounts for the contact part of the 
nucleon-nucleon interaction
\begin{equation}
V_{\delta\pi} = g^\prime \lp( \frac{f_\pi}{m_\pi} \rp)^2 
\vec{\tau}_1\vec{\tau}_2 \bm{\Sigma}_1 \cdot \bm{\Sigma}_2 
\delta (\bm{r}_1-\bm{r}_2)\; ,
\label{deltapi}
\end{equation} 
with the parameter 
$g^{\prime}$ adjusted
to reproduce experimental data on the GTR excitation energies. 

In the $pp$-channel of the RHB model we have used a
phenomenological pairing interaction, the pairing part of the Gogny force,
\begin{equation}
V^{pp}(1,2)~=~\sum_{i=1,2}e^{-[(\mathbf{r}_{1}-\mathbf{r}_{2})/{\mu _{i}}%
]^{2}}\,(W_{i}~+~B_{i}P^{\sigma }-H_{i}P^{\tau }-M_{i}P^{\sigma }P^{\tau }),
\label{Gogny}
\end{equation}
with the set D1S \cite{Ber.84} for the parameters $\mu _{i}$, $W_{i}$, $%
B_{i} $, $H_{i}$ and $M_{i}$ $(i=1,2)$. This force has been very carefully
adjusted to the pairing properties of finite nuclei all over the periodic
table. In particular, the basic advantage of the Gogny force is the finite
range, which automatically guarantees a proper cut-off in momentum space.
For the $T=0$ proton-neutron pairing interaction in open shell nuclei 
we employ a similar interaction which was used in the non-relativistic QRPA 
calculation \cite{Eng.99} of $\beta$-decay rates for spherical neutron-rich
$r$-process waiting-point nuclei.
It consists of a short-range repulsive Gaussian combined with a
weaker longer-range attractive Gaussian
\begin{equation}
\label{eq2}
V_{12}
= - V_0 \sum_{j=1}^2 g_j \; {\rm e}^{-r_{12}^2/\mu_j^2} \;
    \hat\Pi_{S=1,T=0}
\quad ,
\label{pn-pair}
\end{equation}
where $\hat\Pi_{S=1,T=0}$ projects onto states with $S=1$ and $T=0$.  
We follow the prescription from Ref.~\cite{Eng.99}, and take
the ranges $\mu_1$=1.2\,fm and $\mu_2$=0.7\,fm of the two Gaussians from the
Gogny interaction (\ref{Gogny}). The relative strengths $g_1 =1$ and
$g_2 = -2$ are chosen so that the force is repulsive at small distances. 
The overall strength of the force represents the only remaining free parameter. 
We have already successfully applied this force in an analysis of the 
Gamow-Teller resonances in tin isotopes~\cite{Paa.04}.

\subsection{Calculation of $\beta$-decay half-lives}

The rate for the decay of an even-even nucleus in the allowed Gamow-Teller
approximation reads
\begin{equation}
\label{halflife}
\frac{1}{T_{1/2}}=\sum_m{\lambda_{if}^m} = D^{-1}g_A^2\sum_m
  \int{dE_e \lp| \sum_{pn} <1^+_\lambda||\bm{\sigma}\tau_-||0^+>\rp|^2 
   \frac{dn_m}{dE_e}}\;,
\end{equation}
where $D=6163.4\pm 3.8$ s \cite{BG.00}.
$\ket{0^+}$ denotes the ground state of the parent nucleus, and
$\ket{1^+_\lambda}$ is a state of the daughter nucleus.
The sum runs over all final states with an excitation energy smaller than the
$Q_{\beta^-}$ value. In order to account for the universal quenching of
the Gamow-Teller strength function, we use the effective weak axial nucleon
coupling constant $g_A=1$, instead of $g_A=1.26$~\cite{BM.75}. The kinematic
factor in Eq.~(\ref{halflife}) can be written as
\begin{equation}
\label{kinematic}
\frac{dn_m}{dE_e}=E_e\sqrt{E_e^2-m_e^2}(\omega -E_e)^2 F(Z,A,E_e)\;,
\end{equation}
where $\omega$ denotes the energy difference between the initial 
and the final state. The Fermi function $F(Z,A,E_e)$ corrects the phase-space
factor for the nuclear charge and finite nuclear size effects~\cite{KR.65}. 

\subsection{The nucleon effective mass in the RMF models}

In nonrelativistic mean-field models the effective nucleon mass $m^*$
characterizes the energy dependence of an effective local potential which is
equivalent to the nonlocal and frequency dependent microscopic nuclear
potential~\cite{MS.86}. $m^*$ represents a measure
of the density of single-nucleon states around the Fermi surface and,
therefore, it has a pronounced effect on the calculated 
properties of ground and excited states. In the case of
Skyrme-type interactions, for instance, calculation of ground-state
properties and excitation energies of quadrupole giant resonances 
have shown that a realistic choice for the nucleon effective 
mass is in the interval $m^*/m = 0.8\pm 0.1$~\cite{Rei.99}.

In the relativistic mean-field framework the expression ``effective mass"
has been used to denote different quantities. The quantity 
which is usually used to characterize an effective interaction,
and which in the literature is most often called
``the relativistic effective mass", is also known as the 
``Dirac mass"~\cite{JM.89}
\begin{equation}
\label{diracmass}
m_D = m + S(\bm{r})\;,
\end{equation} 
where $m$ is the bare nucleon mass and $S(\bm{r})$ denotes the scalar
nucleon self-energy. The concept of the effective nucleon mass in the
relativistic framework has been extensively analyzed 
in Refs.~\cite{JM.89,JM.90}. Specifically, it has
been pointed out that the Dirac mass should not be identified with the
effective mass of the nonrelativistic mean-field models. Instead, the
quantity which should be compared with the empirical 
effective mass derived from
the nonrelativistic analyses of scattering and bound state 
data is given by
\begin{equation}
\label{effmass}
m^*/m = 1-V/m\;,
\end{equation}
where $V$ denotes the time-like component of the vector self-energy.
The Dirac mass, on the other hand, is determined by two factors: (i) the
empirical spin-orbit splittings in finite nuclei, and (ii) the binding energy at
the saturation density in nuclear matter. In the first order approximation, and
assuming spherical symmetry, the spin-orbit part of the effective single-nucleon
potential reads
 \begin{equation}
\label{soterm}
V_{S.O.}(r)=\frac{1}{4\bar{M}^2}\left( \frac{1}{r}\frac{d}{d r}
       (V-S)\right) \bm{l} \cdot \bm{\sigma}\;,
\end{equation}
where $\bar{M}$ is specified as
\begin{equation}
\bar{M}=M-\frac{1}{2}(V-S)\;.
\end{equation}
While the difference between the vector and scalar potentials determines 
the spin-orbit potential, their sum defines the effective 
single-nucleon potential and is determined by the nuclear matter
binding energy at saturation density. The energy spacings between 
spin-orbit partner states in finite nuclei, and the nuclear matter
binding and saturation, place the following
constraints on the values of the Dirac mass and the nucleon
effective mass: $0.55 m \le m_D \le 0.6 m$, 
$0.64 m \le m^* \le 0.67 m$, respectively. These values 
have been used in most standard relativistic mean-field effective 
interactions. First, we notice that in
comparison to the nonrelativistic models, 
the relativistic nucleon effective mass has a rather
low value, and this results in a smaller density of states 
around the Fermi surface. Second, the range of allowed values of the 
nucleon effective mass is very narrow in the standard relativistic 
mean-field phenomenology, and 
there is really no room for any significant enhancement of the 
single-nucleon level densities at the Fermi surface.

In Ref.~\cite{VNR.02} we have extended the standard relativistic mean-field
model by including dynamical effects originating from the coupling of
single-nucleon motion to collective surface vibrations. 
By using a simple linear
ansatz for the energy dependence of the scalar and 
vector nucleon self-energies,
we were able to describe simultaneously bulk nuclear properties
and single-nucleon spectra in a self-consistent relativistic framework.

As we have emphasized in the introduction, in order to be able to 
reproduce the data on $\beta$-decay lifetimes, 
the description of single-particle energies around the Fermi surface
has to be improved. In principle one could use the method 
of Ref.~\cite{VNR.02} with energy-dependent nucleon self-energies, 
but here we will try a simpler approach to increase the density of 
single-nucleon states without going beyond mean-field level. 
An increase of the effective mass necessitates a reduction of 
the vector self-energy (see Eq. (\ref{effmass})). However, 
in order to retain the empirical value of the
nuclear matter binding energy, the 
scalar self-energy should be reduced correspondingly. 
A serious problem arises
because such an effective interaction 
would systematically underestimate the spin-orbit
splittings in finite nuclei. A solution to this problem has been  
known for a long time, namely the tensor coupling of the 
$\omega$-meson to the nucleon. This interaction enhances 
the effective spin-orbit potential in finite nuclei, but 
is not included in the most commonly used relativistic 
mean-field models. In 
Ref.~\cite{FRS.98} it was shown that the tensor coupling
\begin{equation}
\label{tensor}
\mathcal{L}_{tensor}=-\frac{f_V}{2M}\bar{\psi}\sigma^{\mu\nu}\psi 
(\partial_\mu\omega_\nu - \partial_\nu \omega_\mu )\;,
\end{equation}
generates an additional term in the spin-orbit part of the effective nucleon
potential, which now reads
\begin{equation}
\label{soterm_new}
V_{S.O.}(r)=\left[ \frac{1}{4\bar{M}^2} \frac{1}{r}\frac{d}{d r}
       (V-S) +\frac{f_V}{2M\bar{M}}\frac{1}{r}\frac{d\omega}{d r} \right]
       \bm{l} \cdot \bm{\sigma}\;.
\end{equation}
With the inclusion of the tensor omega-nucleon coupling 
it becomes possible to reproduce the empirical
spin-orbit splittings, even when using effective interactions 
with the Dirac mass as large as $m_D\approx 0.7 m$.

Our strategy is now to use the additional tensor-coupling term 
to generate a density-dependent relativistic effective interaction, 
with a value for the effective mass close to those used in 
nonrelativistic mean-field models.
Starting from the DD-ME1 interaction that we have used in the PN-RQRPA
analysis of charge-exchange modes \cite{Paa.04}, and 
with the inclusion of the additional tensor 
omega-nucleon interaction (\ref{tensor}),
the parameters of the new interaction have been adjusted simultaneously to
properties of nuclear matter and finite nuclei~\cite{NVFR.02}. An
additional constraint 
has been placed on the value of the nucleon effective mass.
The modified effective interaction, denoted as DD-ME1*, exhibits the following
values for the Dirac mass and the nucleon effective mass:
$m_D = 0.67m$, $m^* = 0.76 m$, respectively. These are the highest 
values for which a realistic description of nuclear matter and finite 
nuclei is still possible, i.e. the quality of the calculated 
nuclear matter equation of state and of ground-state properties of spherical 
nuclei is comparable to that of the DD-ME1 interaction. The value 
of the Dirac mass is also in agreement with the results of Ref.~\cite{FRS.98},
where a detailed analysis was performed on the correlation between the 
isoscalar tensor coupling and the Dirac mass in successful mean-field models.  
Although the value of $m^*$ is still lower than those typically used
in nonrelativistic mean-field models, this result presents a significant
improvement over the standard DD-ME1 
density-dependent interaction ($m_D = 0.58 m$, $m^*=0.66 m$).

We have used the new interaction to calculate 
the energy spacings between spin-orbit partner
states in the doubly closed-shell nuclei $^{16}$O, $^{40}$Ca, 
$^{48}$Ca, $^{132}$Sn, and
$^{208}$Pb. The results are shown in Table~\ref{TabA}, 
in comparison with those
obtained using the DD-ME1 interaction, and with the experimental data.
Both interactions provide an excellent description of the
spin-orbit splittings in finite nuclei. In order to illustrate the 
effect of the tensor-coupling term, in Fig.~\ref{Fig1} we display the
radial dependence of the spin-orbit term
of the single-nucleon potential in the self-consistent solutions for the 
ground-state of $^{132}$Sn, calculated with the DD-ME1 and DD-ME1*
effective interactions. For DD-ME1*, in particular, also the contributions
of the first and second term in Eq. (\ref{soterm_new}) are plotted  
separately. We notice that even though the strength of the 
spin-orbit interaction that arises from the large
scalar and vector self-energies (first term in Eq. (\ref{soterm_new}))
is significantly reduced, the tensor omega-nucleon coupling effectively
compensates this reduction, and the resulting spin-orbit potential is
even slightly stronger than the one calculated with the DD-ME1 interaction.
Therefore, while both interactions produce very similar results 
for the spin-orbit splittings in finite nuclei, the inclusion of the 
isoscalar tensor-coupling term in DD-ME1* allows for an increase 
of the Dirac mass and effective mass.  

In Figs.~\ref{Fig2} and \ref{Fig3} we display the neutron and proton 
single-particle levels in $^{78}$Ni and $^{132}$Sn, respectively, 
calculated with DD-ME1, DD-ME1*, and with two nonrelativistic
Skyrme interactions that have been used in the calculation of 
$\beta$-decay half-lives: SkO'~\cite{Eng.99} and SkSC17~\cite{BG.00}. 
Both nonrelativistic interactions are characterized by large
values of the nucleon effective mass: $m^*=0.89$ for SkO', and $m^*=1.0$ for
SkSC17. Compared to the original DD-ME1 interaction, the enhancement 
of the effective mass in DD-ME1* results in the increase of the density
of states around the Fermi surface, similar to the spectra calculated 
with the nonrelativistic interactions.
In Table~\ref{TabB} we compare the calculated neutron and proton 
single-particle levels in $^{132}$Sn, with experimental data.
The levels calculated with DD-ME1* are in much better agreement with 
data~\cite{Isa.02}, than those obtained with the original DD-ME1 interaction.

In Fig.~\ref{Fig4} we plot the evolution of 
the $\pi 2p_{3/2}$ and $\pi 1f_{5/2}$
single-particle levels in heavy copper isotopes. 
The level structure has been
studied in a recent experiment~\cite{Fra.98}, and it has been pointed out
that a possible reordering of these proton levels could have a 
pronounced effect on the shell structure and decay properties 
for nuclei near and beyond the
doubly magic $^{78}$Ni. The results calculated with the DD-ME1* 
interaction are in excellent agreement with the
experimental data. The model reproduces the observed lowering of 
the $\pi 1f_{5/2}$ state
in $^{69}$Cu, $^{71}$Cu, and $^{73}$Cu. For the heavier isotopes 
experimental data are not yet 
avaliable. Nevertheless, our calculation predicts a further 
lowering of the $\pi 1f_{5/2}$ state,
and already in $^{75}$Cu isotope the $\pi 1f_{5/2}$ state becomes 
the ground state.
\section{\label{secIII} $\beta$-decay half-lives}

In Ref.~\cite{Paa.04} we have employed the proton-neutron RQRPA 
in a study of the high-energy part of the GT strength function, 
in particular the GT resonance which represents a coherent 
superposition of high-lying
proton-particle -- neutron-hole configurations associated 
with charge-exchange excitations from $j=l+\frac{1}{2}$ neutron 
orbitals into $j=l-\frac{1}{2}$ proton orbitals.
In addition to the high-energy resonance, the GT
strength function displays a concentration of strength in the 
region of low-energy excitations.
These transitions correspond to core-polarization 
($j=l\pm\frac{1}{2} \to j=l\pm\frac{1}{2}$), and back spin-flip 
($j=l-\frac{1}{2} \to j=l+\frac{1}{2}$) neutron-hole -- proton-particle
excitations. It is precisely this low-energy tail that contributes to
the $\beta$-decay process.

Only the isovector channel of the effective interaction contributes to the
matrix elements of the residual interaction in 
the calculation of charge-exchange excitations. Nevertheless,  
the isoscalar channel plays an important role because it determines
the single-particle levels that enter the PN-RQRPA calculations. 
In the analysis of charge-exchange modes we have used the DD-ME1 effective
interaction, combined with the standard parameters for the
pion-nucleon Lagrangian: $m_\pi = 138$ MeV, and $f_\pi^2/4\pi = 0.08$. The
parameter of the zero-range Landau-Migdal force $g^\prime = 0.55$ has been 
adjusted to reproduce the excitation energy of the GT resonance in $^{208}$Pb.
We have shown that, although the inclusion of $T=0$ pairing in the
residual interaction does affect both the low- and high-energy 
regions of the GT strength distribution, 
the position of the resonance is not sensitive to its strength~\cite{Paa.04}. 
On the other hand, $T=0$ pairing has a very strong
influence on the low-lying tail of the GT strength distribution.

Our first attempt to apply the PN-RQRPA to $\beta$-decay processes, 
by employing the DD-ME1 interaction, was not
successful. We tried to reproduce the empirical half-life of the
$^{78}$Ni. This spherical nucleus undergoes high-energy fast $\beta$-decay
and, since it has doubly-closed spherical shells, 
there is no contribution from the $T=0$ pairing. The PN-RQRPA with
the DD-ME1 interaction predicts a 
half-life $T_{1/2}=7$ s, which is an order of magnitude
longer than the experimental value $T_{1/2}=104+126-57$ ms~\cite{Hosmer}. The 
results of 
nonrelativistic QRPA calculations are much closer to the empirical half-life.
For example, for the interaction DF3 
the calculated half-life is $T_{1/2}\approx 0.3$ s~\cite{Bor.03}, 
whereas the Skyrme interaction SkO' predicts 
$T_{1/2}\approx 0.6$ s~\cite{Eng.99}.
Obviously, while the DD-ME1 interaction provides an accurate description 
of the high-energy region of the GT strength function, it does a rather
poor job for the low-energy tail of the distribution. 
The problem, as we have already emphasized, lies in the
low density of single-proton states around the Fermi level. This has motivated
the adjustment of the new effective interaction
DD-ME1*, which has a considerably higher value of the nucleon effective mass, 
and consequently produces a higher density of single-nucleon states at 
the Fermi surface.

The change of the density of single-nucleon states will also affect 
the position of the GT resonance, and therefore the parameter 
of the zero-range Landau-Migdal force 
has to be readjusted. A higher density of states implies a lowering 
of the GT strength distribution, and this means that value of 
$g^\prime = 0.55$ used with DD-ME1, has to be increased in order 
to reproduce the empirical excitation energy of GT resonances. 
For DD-ME1* we have adjusted the new value $g^\prime = 0.62$ to
the position of the GT resonance in $^{208}$Pb. With this 
value of $g^\prime$ we also find an excellent agreement between 
the calculated and experimental excitation energies of the GTR for
$^{48}$Ca, $^{90}$Zr, and $^{112-124}$Sn.

With the new set of parameters (DD-ME1* effective interaction, 
$m_\pi = 138$ MeV, $f_\pi^2/4\pi = 0.08$, $g^\prime = 0.62$), 
we have recalculated the $\beta$-decay
half-life of $^{78}$Ni. The value $T_{1/2}=0.9$ s presents a
significant improvement over our first result obtained with DD-ME1, 
although it still overestimates the empirical
half-life of $^{78}$Ni, and the values calculated with the nonrelativistic
PN-QRPA. The most probable reason is that the value of the nucleon 
effective mass used in DD-ME1* 
is still below the values used in the nonrelativistic effective
interactions \cite{Eng.99,Bor.03}. A further increase of $m^*$ in 
our relativistic model is, however, not possible without downgrading 
the agreement with experimental data on ground state properties 
of finite nuclei.

In recent years a number of experimental and theoretical
studies have focused on the level structure and decay properties of
neutron-rich nuclei in the vicinity of the doubly magic $^{78}$Ni and 
$^{132}$Sn. In particular, beta-decay rates 
in these two regions of the periodic chart
have been extensively investigated in the framework of the nonrelativistic
PN-QRPA~\cite{Eng.99,BG.00,Bor.03}. In the next two sections we apply 
the newly formulated relativistic PN-QRPA in the calculation of 
$\beta$-decay half-lives of nuclei in the regions N$\approx$50 and 
N$\approx$82.

\subsection{N$\approx$50 region}

In this region we have investigated the iron, nickel, and zinc isotopic chains,
and the $N=50$ isotones. The structure of the low-energy part of the GT strength
distribution crucially depends on the occupancy of the $Z=28$ proton
shell. It should be noted that, because of possible deformation effects 
in the Fe and Zn chains, these nuclei might not be as good as the Ni 
isotopes for a comparison with results of spherical PN-QRPA calculations.

In Fig.~\ref{Fig5} we display the calculated half-lives of the Fe isotopes 
(two holes in the $\pi 1f_{7/2}$ orbit) in
comparison with the available experimental data \cite{NUBASE}. 
The results are obtained with the DD-ME1* interaction, 
$m_\pi = 138$ MeV, $f_\pi^2/4\pi = 0.08$, $g^\prime = 0.62$,
Gogny D1S T=1 pairing. Two values for the strength parameter
of the $T=0$ pairing interaction (\ref{eq2}) have been used.
Obviously, the absolute values of the calculated half-lives 
are very sensitive to the $T=0$ pairing strength. Without the inclusion 
of this pairing channel, in all considered cases the calculated 
half-lives are at least an order of magnitude longer than the 
experimental values. In the case of iron isotopes 
the pairing strength parameter $V_0=115$ MeV has been adjusted 
to reproduce the half-life of 
$^{68}$Fe. With this value we are 
able to reproduce the half-lives of 
the $^{66}$Fe and $^{70}$Fe isotopes very 
accurately, whereas the lifetime of the $^{64}$Fe isotope is somewhat
overestimated. 

It is, however, probable that the inclusion of a strong $T=0$ pairing 
partially compensates the deficiencies of the single-particle spectra 
calculated with the DD-ME1* interaction. 
For the semi-magic nucleus $^{76}$Fe,
in Fig.~\ref{Fig6} we plot the probabilities
$\lambda_{fi}$ of $\beta$-transitions from an initial nuclear 
state $i$ to a final state $f$, for different values of the 
$T=0$ pairing strength. 
The ground-state occupation probabilities for the single-particle levels 
relevant for this $\beta$-decay process are listed in Tab.~\ref{TabC}. 
Since the $\pi 1f_{7/2}$ orbit is not fully occupied, 
the transition with the highest probability is dominated (95\% of the
neutron-to-proton QRPA amplitude) by the back spin-flip configuration
$\nu 1f_{5/2} \to \pi 1f_{7/2}$. Other transitions, with much
smaller probabilities, correspond to the back spin-flip configuration
$\nu 2p_{1/2} \to \pi 2p_{3/2}$, and the core-polarization configurations
$\nu 1f_{5/2} \to \pi 1f_{5/2}$, $\nu 1g_{9/2} \to \pi 1g_{9/2}$,
$\nu 2p_{3/2} \to \pi 2p_{3/2}$, and $\nu 2p_{1/2} \to \pi 2p_{1/2}$.
Since except $\pi 1f_{7/2}$,
all proton single-particle levels listed in Tab.~\ref{TabC} have very small 
occupation probabilities, the only sizeable contribution from the 
$T=0$ pairing to the RQRPA matrices comes from the
$\pi 1f_{7/2} (\nu 1f_{5/2})^{-1}$ pair. Because of the attractive 
nature of the pairing interaction, the large diagonal matrix element 
$v_p^2 V_{pnpn}^{pp}$ 
(p and n denote $\pi 1f_{7/2}$ and $\nu 1f_{5/2}$ states, respectively) 
effectively reduces the sum of the quasiparticle energies:
$H_{pp}^{11}+H_{nn}^{11} = E_p + E_n$. This
means that the $T=0$ pairing compensates for the
fact that even the DD-ME1* interaction 
still predicts a rather low density of states
around the Fermi surface, i.e., the  $\pi 1f_{7/2}$ and $\nu 1f_{5/2}$
single-particle levels are still too close. 
The inclusion of the $T=0$ pairing will affect only configurations with the 
proton level at least partially occupied.  This is clearly seen 
in Fig.~\ref{Fig6}, where we notice that only the transition
built from the configuration
$\nu 1f_{5/2} \to \pi 1f_{7/2}$ is lowered in energy
and enhanced with the increase of the strength of $T=0$ pairing, 
whereas the remaining transitions are unaltered.

For the Ni isotopes the $\pi 1f_{7/2}$ orbit is  
completely occupied, i.e., in this case 
the transition $\nu 1f_{5/2} \to \pi 1f_{7/2}$ is blocked.
The $T=0$  pairing could only have an effect on the
$\pi 1g_{9/2} (\nu 1g_{9/2})^{-1}$ configuration, 
because the $\nu 1g_{9/2}$ orbit is not fully occupied for isotopes
below $^{78}$Ni. However, it is true that when this orbit is
almost empty ($^{70}$Ni, $^{72}$Ni) there is a large contribution from the
pairing interaction, but at the same time there is only a small number of
neutrons which can participate in the $\beta$-decay process. 
On the other hand, when this orbit is almost full ($^{74}$Ni, $^{76}$Ni), 
the contribution from the $T=0$
pairing becomes negligible. In contrast to the iron isotopic chain,
we were not able to obtain a single value of the $T=0$ pairing strength 
parameter that would provide a consistent description of the entire 
chain of nickel isotopes. To reproduce the experimental 
half-lives, an extremely strong 
$T=0$ pairing would have to be used and this would cause the 
collapse of the PN-RQRPA calculation. Because of the closed $Z=28$ proton 
shell, the $T=0$ pairing is not effective 
in Ni isotopes and, therefore, the calculated $\beta$-decay half-lives,
shown in Fig.~\ref{Fig7}, overestimate the experimental data.

In this mass region we have also analyzed the half-lives of zinc isotopes. 
In Ref.~~\cite{Eng.99} the measured half-lives of three zinc isotopes 
$^{76}$Zn, $^{78}$Zn, and $^{80}$Zn, and of $^{82}$Ge, were used to 
adjust the $T=0$ pairing strength. It was shown that the experimental 
lifetimes can all be reproduced with a single value of the  
$T=0$ pairing strength, $V_0 = 230$ MeV. In the present calculation
the measured half-lives are only reproduced by using a much stronger 
$T=0$ pairing interaction. The effect that $T=0$ pairing has on the 
$\beta$-decay probability is illustrated in the example of 
the semi-magic nucleus $^{80}$Zn.
In Tab.~\ref{TabC} the ground-state occupation probabilities 
for the relevant single-particle levels are included.
It is important to note that, because of the $T=1$ pairing, 
the occupation probability of the
$\pi 1f_{7/2}$ state is not equal to one. Besides $\pi 1f_{7/2}$, 
only $\pi 1f_{5/2}$
has a sizeable occupation probability among the proton states. 
The $T=0$ pairing interaction
produces a large contribution to the RQRPA matrices for the following
configurations: $\pi 1f_{5/2}(\nu 1f_{5/2})^{-1}$ and 
$\pi 1f_{7/2}(\nu 1f_{5/2})^{-1}$. Because the transition 
$\nu 1f_{5/2} \to \pi 1f_{7/2}$ is essentially blocked, the effect of
$T=0$ pairing is much weaker than in the case of Fe isotopes.
If the parameter $V_0$ is kept below $\approx 250$ MeV, the $T=0$ pairing has
virtually no effect on the calculated half-lives. In the interval
between $V_0 = 0$ MeV and $V_0 = 220$ MeV the half-life decreases 
by just 10\%. With a further increase of $V_0$, however, the calculated  
half-life displays a steep decrease. The measured half-life is 
reproduced for $V_0 \approx 330$ MeV.
The corresponding distributions of neutron-to-proton QRPA amplitudes 
for three values of the
$T=0$ pairing strength ($V_0=0$ MeV, $V_0=255$ MeV, and $V_0=330$ MeV) are 
shown in Tab.~\ref{TabD}. 
As one would expect, in the absence of $T=0$ pairing,
the $\beta$-decay process is characterized by two transitions. 
The lower one is dominated by the back spin-flip transition
$\nu 2p_{1/2} \to \pi 2p_{3/2}$, whereas the higher component 
represents a mixture of core-polarization transitions.
Increasing the $T=0$ pairing strength to $V_0=255$ MeV, we note that:
(i) transitions built from the back spin-flip configuration 
$\nu 2p_{1/2} \to \pi 2p_{3/2}$ and the core-polarization 
configurations are no
longer well separated, and (ii) an additional transition appears, built
dominantly on the back spin-flip configuration 
$\nu 1f_{5/2} \to \pi 1f_{7/2}$, which  
results in a sudden reduction
of the calculated half-life. Of course, this 
would not be possible if $\pi 1f_{7/2}$ was fully occupied.
This was the case for the Ni isotopes, and consequently their
half-lives could not be improved by
increasing the strength of the $T=0$ pairing. 
Since the $T=0$ pairing has a strong effect on the 
$\nu 1f_{5/2} \to \pi 1f_{7/2}$ configuration, 
a further increase of its strength
lowers the energy of the corresponding transition.
For $V_0=330$ MeV we find only one transition, predominantly based on 
the $\nu 1f_{5/2} \to \pi 1f_{7/2}$ configuration.

Although $V_0=330$ MeV reproduces the experimental half-lives of Zn
isotopes, as shown in Fig.~\ref{Fig8}, this does not mean that one 
should use this particular
value  for the other isotopic chains, even in the same mass region.
This is illustrated in Fig.~\ref{Fig9}, where we display the half-lives of the
$N=50$ isotones for three values of the $V_0$ parameter: $V_0=0$ MeV,
$V_0=115$ MeV (used in the calculation of Fe isotopes), 
and $V_0=330$ MeV (used for Zn isotopes). In contrast to the result 
of Ref.~\cite{Eng.99}, the value $V_0=330$ MeV which reproduces 
the half-lives of Zn nuclei, overestimates the half-life
of $^{82}$Ge. In order to reproduce the
experimental value $T_{1/2}=4.55\pm0.05$ s~\cite{NUBASE}, the pairing strength 
must be increased to $V_0=345$ MeV. Both values, $V_0=330$ MeV and 
$V_0=345$ MeV, are rather close to the point at which the PN-RQRPA 
calculation collapses. This is illustrated in 
Fig.~\ref{Fig10}, where we plot the half-lives of 
$^{78}$Zn, $^{80}$Zn, $^{82}$Zn, and $^{82}$Ge as functions of the 
$T=0$ pairing strength parameter. The PN-RQRPA instability point is, 
for all four nuclei, located at $V_0 \approx 400$ MeV. 

\subsection{N$\approx$82 region}

In the mass region around the doubly magic $^{132}$Sn, 
we have calculated the $\beta$-decay half-lives of
the cadmium, tin, and tellurium isotopic chains, as well as the
$N=82$ isotones. In Fig.~\ref{Fig11} we plot the calculated half-lives of the
Cd isotopes. The results correspond to two calculations, with  
$V_0=0$ and $V_0=225$ MeV for the strength parameter of the $T=0$
pairing. As in the cases that we have considered in the previous 
section, the calculated half-lives are more than an order of magnitude
too long when the $T=0$ pairing is not included. With the 
pairing strength parameter $V_0=225$ 
MeV adjusted to reproduce the half-life of $^{130}$Cd, 
the PN-QRPA calculation reproduces the experimental half-lives of the 
Cd isotopic chain.
With two holes in the $\pi 1g_{9/2}$ orbit, the situation in the 
cadmium chain is similar to that of
Fe isotopes (two holes in the $\pi 1f_{7/2}$ orbit). The $\beta$-decay
process in the Cd isotopes is dominated by the back spin-flip transition 
$\nu 1g_{7/2} \to \pi 1g_{9/2}$. Again, an increase
of the $T=0$ pairing strength partially compensates for the fact that the
difference between the $\nu 1g_{7/2}$ and $\pi 1g_{9/2}$ single-particle
energies is too small, due to a relatively small effective mass.

For the Sn isotopes the $\pi 1g_{9/2}$ orbit is completely occupied and the 
transition $\nu 1g_{7/2} \to \pi 1g_{9/2}$ is blocked. A similar problem
we already encountered for the Ni isotopes. One would, therefore,
expect that the calculated half-lives of the Sn isotopes will overestimate
the experimental values by at least an order of magnitude, and furthermore that
it will not be possible to improve the results by simply 
increasing the $T=0$ pairing
strength. However, this turns out to be true only for $^{132}$Sn.
Our model predicts that this isotope is stable against $\beta$-decay, 
whereas the experimental half-life is
$T_{1/2}=39.7\pm 0.5$ s~\cite{NUBASE}. In the left panel of Fig.~\ref{Fig12} we
display the calculated half-lives of the Sn isotopes, in comparison with the
available experimental data~\cite{NUBASE}. We notice that, in contrast to
$^{132}$Sn, the theoretical half-lives of the heavier Sn isotopes 
show a pronounced dependence on the $T=0$ pairing strength. For $V_0=225$ MeV
the calculated half-lives are only slightly longer than the experimental 
values. 
This is easily explained by the fact that in tin isotopes beyond $^{132}$Sn
neutrons begin to occupy  the $\nu 1h_{9/2}$ single-particle level, and this
enables the back spin-flip transition $\nu 1h_{9/2} \to \pi 1h_{11/2}$.
The occupation probabilities of the $\nu 1h_{9/2}$ single-particle 
level for the Sn and Te isotopes are included in Tab.~\ref{TabE}.
Because the neutron level $\nu 1h_{9/2}$ has low occupancy, the $T=0$ pairing
produces a very strong effect on the $\pi 1h_{11/2}(\nu 1h_{9/2})^{-1}$
configuration, and reduces the calculated half-lives to the experimental
values.

This effect is further illustrated in the middle panel of Fig.~\ref{Fig12}, 
where we plot the calculated half-lives of the Te isotopes, in comparison with
avaliable experimental data. In this case, the choice $V_0=225$ MeV results 
in half-lives that are even somewhat shorter than the experimental values.
This is related to the fact that the calculated energy spacing between the
$\nu 1h_{9/2}$ and $\pi 1h_{11/2}$ states in $^{132}$Sn 
is larger than the experimental value (see Table \ref{TabB}). 
A slightly lower value of the $T=0$ pairing strength ($V_0 = 200$ MeV)  
accurately reproduces the half-lives of the Te isotopes.
Finally, in the right panel of Fig.~\ref{Fig12} we display the calculated
half-lives of the $N=82$ isotones, compared to the shell-model 
results of Ref.~\cite{Lan.99}. This comparison also shows that 
$V_0=225$ MeV presents a realistic choice for the strength of the $T=0$ 
pairing in this mass region.


\section{\label{secIV}Concluding remarks and outlook}

In this work the newly developed relativistic PN-QRPA has been 
for the first time applied to Gamow-Teller $\beta$-decays of 
nuclei relevant for the $r$-process. 
The matrix equations of the QRPA are formulated in the canonical
single-nucleon basis of the relativistic Hartree-Bogoliubov (RHB) model.
The RHB+QRPA model employed in this work is fully self-consistent.
For the interaction in the particle-hole channel effective Lagrangians with
density-dependent meson-nucleon couplings are used, 
and pairing correlations are
described by the pairing part of the finite range Gogny interaction. Both in
the $ph$ and $pp$ channels, the same interactions are used in the RHB
equations that determine the canonical quasiparticle basis, and in the
matrix equations of the RQRPA. The model also includes a proton-neutron
particle-particle interaction ($T=0$ pairing), and here we have employed 
a sum of two Gaussian functions that was used in the non-relativistic QRPA 
calculation \cite{Eng.99} of $\beta$-decay rates for spherical neutron-rich
$r$-process waiting-point nuclei.

Microscopic global predictions of weak interaction rates are very 
important, because most of the neutron-rich nuclei relevant for 
the $r$-process nucleosynthesis are not accessible in experiment. 
Calculated $\beta$-decay half-lives depend on a detailed description
of transition energies, shell structure, and on the choice of the
residual interactions in the $ph$ and $pp$ channels. In the present 
analysis we have shown that standard relativistic mean-field 
effective interactions, adjusted to nuclear matter and ground-state 
properties of spherical nuclei, generally overestimate the empirical 
half-lives by more than an order of magnitude. The main reason is 
their low effective nucleon mass which, in the standard choice of 
interaction terms, is strongly related to the empirical energy 
spacings of spin-orbit partner states. Thus, in order to be able
to reproduce measured $\beta$-decay half-lives, we have to increase
the effective nucleon mass of the relativistic mean-interaction
used in the RHB calculation of nuclear ground state and in the 
$ph$ channel of the QRPA residual interaction. In this work 
we have done it on the mean-field level, by including an 
additional isoscalar tensor-coupling term in the Lagrangian, 
which allows for an increase of the Dirac mass and effective mass, 
while at the same time the new effective interaction reproduces 
the ground-state properties of finite nuclei, including the 
spin-orbit splittings. The new force has been adjusted starting from
our most successful parameter set so far, the density-dependent 
interaction DD-ME1. We note, however, that the inclusion of the 
tensor term allows only for a moderate increase of the effective
mass. We cannot, like in Skyrme forces for example, adjust interactions
with values of the effective mass close to one.

With the new density-dependent interaction DD-ME1* we have calculated the
GT distribution strengths and $\beta$-decay rates of neutron-rich nuclei
in the mass regions N$\approx$50 and N$\approx$82. The model reproduces
in detail the data on GT resonances and the low-energy GT strength 
(see also Ref.~\cite{Paa.04}). The results for 
$\beta$-decay half-lives are similar to those obtained in the 
nonrelativistic PN-QRPA calculation of Ref.~\cite{Eng.99}, where it 
was shown that a fine tuning of the strength of the $T=0$ pairing
interaction is necessary in order to reproduce the experimental 
data. In general, by adjusting the strength parameter of the 
proton-neutron pairing interaction to one experimental half-life, 
the PN-RQRPA calculation reproduces the data for a chain of isotopes. 
In the region N$\approx$50 very different values of the $T=0$ pairing
strength reproduce the empirical lifetimes of the Fe and Zn 
isotopic chains, whereas a single value, adjusted to the half-life
of $^{130}$Cd, qualitatively reproduces the data in the N$\approx$82
region. $T=0$ pairing, however, does not help in the case of Ni 
isotopes and for $^{132}$Sn, and the model overestimates 
the half-lives on Ni nuclei and predicts a $\beta$-stable
$^{132}$Sn. Therefore, we have not been able, as suggested in
Ref.~\cite{Eng.99}, to adjust the strength of the $T=0$ pairing 
on experimental data, and extend the calculation to other mass 
regions, or even to other isotopic chains in the same mass region.
This problem could be due, however, to the deficient description
of shell structure, related to the low effective mass of our
$ph$ interaction. This is also the reason why we did not attempt
a calculation of $\beta$-decay rates in the region N$\approx$126,
where there are not enough data to constrain the strength 
of the $T=0$ pairing interaction.  

The relativistic PN-QRPA nevertheless presents a valuable tool for 
the investigation of weak interaction rates of neutron-rich nuclei.
It can be used for microscopic calculations not only of Gamow-Teller,
but also of first-forbidden decays, which play an important 
role for $\beta$-decay rates both in the N$\approx$82 and N$\approx$126 
regions \cite{Bor.03,She.02}.
In order to be able to make reliable predictions, however, the 
model will have to be improved by allowing for higher effective 
nucleon masses, possibly by going beyond the simple mean-field 
approximation.   

\bigskip \bigskip

\leftline{\bf ACKNOWLEDGMENTS}
\noindent
This work has been supported in part by the Bundesministerium
f\"ur Bildung und Forschung under project 06 TM 193, and by the
Gesellschaft f\" ur Schwerionenforschung (GSI) Darmstadt.
N.P. acknowledges support from the Deutsche
Forschungsgemeinschaft (DFG) under contract SFB 634.
==========================================================================

\newpage
\begin{figure}
\caption{Radial dependence of the spin-orbit term
of the single-nucleon potential in self-consistent solutions for the 
ground-state of $^{132}$Sn. The curves correspond to
RMF calculations with the DD-ME1 and DD-ME1* effective
interactions. For the latter the contributions
of the first and second term in 
Eq.~(\protect\ref{soterm_new}) are plotted  
separately.}
\label{Fig1}
\end{figure}

\begin{figure}
\caption{Neutron and proton 
single-particle levels in $^{78}$Ni 
calculated with DD-ME1 (a), DD-ME1* (b), and with two nonrelativistic
Skyrme interactions: SkO'~\protect\cite{Eng.99} (c) 
and SkSC17~\protect\cite{BG.00} (d).}
\label{Fig2}
\end{figure}

\begin{figure}
\caption{Same as in Fig.~\protect\ref{Fig2}, for $^{132}$Sn.}
\label{Fig3}
\end{figure}

\begin{figure} 
\caption{$\pi 2p_{3/2}$ and $\pi 1f_{5/2}$
single-particle levels in heavy copper isotopes, 
calculated with the DD-ME1* interaction, in 
comparison with experimental data \protect\cite{Fra.98}.}
\label{Fig4}
\end{figure}

\begin{figure} 
\caption{Comparison of the calculated half-lives of Fe isotopes, 
for two values of the $T=0$ pairing strength, with available 
experimental data \protect\cite{NUBASE}.}
\label{Fig5}
\end{figure}

\begin{figure} 
\caption{Probabilities of $\beta$-transitions for the 
semimagic  $^{76}$Fe, for different values of the 
$T=0$ pairing strength.}
\label{Fig6}
\end{figure}

\begin{figure} 
\caption{Calculated half-lives of Ni isotopes, in comparison 
with experimental values. The data are from Ref.\protect\cite{NUBASE},
except for $^{78}$Ni where the value 
$T_{1/2}=104+126-57$ ms \protect\cite{Hosmer} is used.}
\label{Fig7}
\end{figure}

\begin{figure} 
\caption{Calculated half-lives of Zn isotopes, without and 
with the inclusion of $T=0$ pairing, in comparison 
with experimental and extrapolated values\protect\cite{NUBASE}.}
\label{Fig8}
\end{figure}

\begin{figure} 
\caption{Calculated half-lives of $N=50$ isotones for three 
values of the $T=0$ pairing strength. The data are from 
Ref.~\protect\cite{NUBASE}, except for $^{78}$Ni where the value 
$T_{1/2}=104+126-57$ ms \protect\cite{Hosmer} is used.}
\label{Fig9}
\end{figure}

\begin{figure} 
\caption{Calculated half-lives of 
$^{78}$Zn, $^{80}$Zn, $^{82}$Zn, and $^{82}$Ge as functions 
of the $T=0$ pairing strength.}
\label{Fig10}
\end{figure}

\begin{figure} 
\caption{Calculated half-lives of Cd isotopes for two values 
values of the $T=0$ pairing strength, in comparison with 
available experimental data~\protect\cite{NUBASE}.}
\label{Fig11}
\end{figure}

\begin{figure} 
\caption{Calculated half-lives of Sn (left panel) and Te (middle panel)
isotopes for two values of the $T=0$ pairing strength,
in comparison with experimental data~\protect\cite{NUBASE}. 
In the right panel
the results for the $N=82$ isotones are compared with the shell-model
results~\protect\cite{Lan.99}.}
\label{Fig12}
\end{figure}

\newpage

\begin {table}[]
\begin {center}
\caption {Energy separation (in MeV) between spin-orbit partner states
in doubly closed-shell nuclei, calculated with the DD-ME1 and DD-ME1*
interactions, and compared with experimental data~\protect\cite{NUDAT}.}
\begin {tabular}{cccccc}
\hline
\hline
    & &{ DD-ME1}& { DD-ME1*}& &{Exp.}   \\ 
\hline
$^{16}$O    & $\nu 1p$   &  6.32 &  6.02  & & 6.18    \\
            & $\pi 1p$   &  6.25 &  5.96  & & 6.32    \\ 
\hline
$^{40}$Ca   & $\nu 1d$   &  6.57 &  6.59  & & 6.00    \\
            & $\pi 1d$   &  6.51 &  6.51  & & 6.00    \\ 
\hline
$^{48}$Ca   & $\nu 1f$   &  7.69 &  7.79  & & 8.38    \\
            & $\nu 2d$   &  1.72 &  0.56  & & 2.02   \\ 
\hline
$^{132}$Sn  & $\nu 2d$   &  1.88 &  2.03  & & 1.65  \\
            & $\pi 1g$   &  6.24 &  6.57  & & 6.08  \\
        & $\pi 2d$   &  1.82 &  1.98  & & 1.75  \\ 
\hline
$^{208}$Pb  & $\nu 2f$   &  2.20 &  2.38  & & 1.77  \\
            & $\nu 1i$   &  6.84 &  7.13  & & 5.84  \\
        & $\nu 3p$   &  0.88 &  0.89  & & 0.90 \\
        & $\pi 2d$   &  1.65 &  1.82  & & 1.33 \\
        & $\pi 1h$   &  5.84 &  6.06  & & 5.56\\
\hline
\hline
\end{tabular}
\label{TabA}
\end{center}
\end{table}

\begin {table}[]
\caption{Neutron (left panel) and proton (right panel) single-particle
energies $E_{nlj}$ in $^{132}$Sn. For each panel, the left-hand column 
specifies the radial, orbital and total angular momentum
quantum numbers, the column labeled DD-ME1 contains results calculated
with the standard density-dependent RMF interaction DD-ME1, the spectrum
calculated with the new effective interaction DD-ME1* is included
in the third column. The theoretical spectra are compared to the experimental
energies shown in the column labeled EXP.}

\begin{center}
\begin{tabular}{c c c c c c c c c}
\hline
\hline
\multicolumn{4}{c}{\sc neutron states}  & &
\multicolumn{4}{c}{\sc proton states} \\    
\hline 
{$nlj$}	         & {\sc DD-ME1}      & {\sc DD-ME1*}     & {\sc EXP}  &
& {$nlj$}	 & {\sc DD-ME1}      & {\sc DD-ME1*}     & {\sc EXP}  \\ 
\hline
\hline
{$1g_{9/2}$}     & {-18.76}    & {-17.44}   & {} &
& {$2p_{3/2}$}   & {-18.91}   & {-18.79}   & {}  \\ \hline
{$2d_{5/2}$}     & {-10.91}   & {-10.32}   & {-9.04} & 
& {$2p_{1/2}$}    & {-17.62}   & {-17.44}   & {-16.13} \\ \hline
{$1g_{7/2}$}     & {-12.86}   & {-11.20}   & {-9.82} & 
& {$1g_{9/2}$}     & {-16.25}   & {-16.55}   & {-15.78} \\ \hline
{$3s_{1/2}$}     & {-8.68}   & {-8.12}   & {-7.72} & 
& {$1g_{7/2}$}     & {-10.04}   & {-9.98}   & {-9.65}  \\ \hline
{$1h_{11/2}$}     & {-7.42}   & {-7.30}   & {-7.63} & 
& {$2d_{5/2}$}     & {-7.60}   & {-8.58}  & {-8.69} \\ \hline
{$2d_{3/2}$}     & {-8.96}   & {-8.29}   & {-7.39}  &
& {$2d_{3/2}$}     & {-5.70}  & {-6.60}  & {-6.95} \\ \hline
{$2f_{7/2}$}    & {-1.40}   & {-1.29}   & {-2.45}  &
& {$1h_{11/2}$}     & {-5.30}  & {-6.78}  & {-6.86} \\ \hline
{$3p_{3/2}$}     & {-0.58}   & {-0.35}   & {-1.59}  &
& {$3s_{1/2}$}     & {-4.88}  & {-5.87}  & {} \\ \hline
{$1h_{9/2}$}     & {-0.11}   & {0.45}   & {-0.88} & 
& {$2f_{7/2}$}     & {2.66}  & {1.32}  & {} \\ \hline
{$3p_{1/2}$}    & {-0.16}   & {0.05}   & {-0.75} & 
&            &          &          &       \\ \hline
{$2f_{5/2}$}     & {0.38}   & {0.56}   & {-0.44} &
&           &         &        &         \\ \hline \hline
\end{tabular}
\end{center}
\label{TabB}
\end{table}

\begin {table}[h]
\caption{Occupation probabilities of neutron and proton single-particle
states for the ground-states of $^{76}$Fe and $^{80}$Zn.}
\begin {center}
\begin {tabular}{c c c c c c c}
\hline
\hline
\multicolumn{3}{c}{$^{76}$Fe} & & \multicolumn{3}{c}{$^{80}$Zn} \\ \hline
{$nlj$}	&{\sc neutrons} & {\sc protons} & & {$nlj$} & {\sc neutrons} 
& {\sc protons}    
\\ \hline 
{$1f_{7/2}$} & {1.000}  & {0.743}  & & {$1f_{7/2}$}  & {1.000}  & {0.980}  \\
{$1f_{5/2}$} & {1.000}  & {0.019}  & & {$1f_{5/2}$}  & {1.000}  & {0.260} \\ 
{$2p_{3/2}$} & {1.000}  & {0.005}  & & {$2p_{3/2}$}  & {1.000}  & {0.098} \\ 
{$2p_{1/2}$} & {1.000}  & {0.004}  & & {$2p_{1/2}$}  & {1.000}  & {0.040}  \\ 
{$1g_{9/2}$} & {1.000}  & {0.004}  & & {$1g_{9/2}$} & {1.000}  & {0.014}\\ 
\hline
\hline
\end {tabular}
\end{center}
\label{TabC}
\end{table}

\begin {table}[]
\caption{$\beta^-$-transitions in $^{80}$Zn for three values of 
the $T=0$ pairing strength. The contribution (in percent) of a particular 
configuration to the QRPA amplitude is included in parenthesis. }
\begin{center}
\begin{tabular}{c c c c c c}
\hline
\hline
\multicolumn{2}{c}{$V_0=0$ MeV ($T_{1/2}=18.9$ s)}  & 
\multicolumn{2}{c}{$V_0=255$ MeV ($T_{1/2}=13.1$ s) } &
\multicolumn{2}{c}{$V_0=330$ MeV ($T_{1/2}=0.6$ s) }   \\
\hline
{$E_f-E_i$ (MeV)}  &  & {$E_f-E_i$ (MeV)} &  &  {$E_f-E_i$ (MeV)} \\ 
\hline
{-3.088}  & {$g_{9/2} \to g_{9/2}$ (12\%)} & 
{-3.271}  & {$f_{5/2} \to f_{5/2}$ (11\%)} &
{-5.273}  & {$f_{7/2} \to f_{7/2}$ (5\%)}  \\
{}  & {$p_{1/2} \to p_{3/2}$ (80\%)} & 
{}  & {$g_{9/2} \to g_{9/2}$ (18\%)} &
{}  & {$f_{5/2} \to f_{7/2}$ (71\%)}  \\
{}  & {} & 
{}  & {$p_{1/2} \to p_{3/2}$ (60\%)} &
{}  & {}  \\
{-2.690}  & {$p_{3/2} \to p_{3/2}$ (6\%)} & 
{-2.982}  & {$f_{5/2} \to f_{5/2}$ (13\%)} &
{} & {}  \\
{}  & {$p_{1/2} \to p_{3/2}$ (11\%)} & 
{}  & {$g_{9/2} \to g_{9/2}$ (16\%)} &
{} & {}  \\
{}  & {$f_{5/2} \to f_{5/2}$ (22\%)} & 
{}  & {$p_{1/2} \to p_{3/2}$ (19\%)} &
{} & {}  \\
{}  & {$g_{9/2} \to g_{9/2}$ (54\%)} & 
{}  & {$p_{3/2} \to f_{5/2}$ (47\%)} &
{} & {}  \\
{}  & {} & 
{-2.302}  & {$p_{1/2} \to p_{1/2}$ (13\%)} &
{} & {}  \\
{}  & {} & 
{}  & {$g_{9/2} \to g_{9/2}$ (15\%)} &
{} & {}  \\
{}  & {} & 
{}  & {$f_{5/2} \to f_{7/2}$ (58\%)} &
{} & {}  \\
\hline
\hline
\end {tabular}
\end{center}
\label{TabD}
\end{table}

\begin {table}[h]
\caption{Occupation probabilities of the $\nu 1h_{9/2}$ single-particle
state for the ground-states of the Sn and Te isotopes.}
\begin {center}
\begin {tabular}{c c c c c}
\hline
\hline
\multicolumn{2}{c}{Sn} & & \multicolumn{2}{c}{Te} \\ \hline
{A}	&{$v_{\nu 1h_{9/2}}^2$} & & {A} & {$v_{\nu 1h_{9/2}}^2$} 
\\ \hline 
{134} & {0.024}   & & {136}  & {0.035}    \\
{136} & {0.054}   & & {138}  & {0.077}   \\ 
{138} & {0.092}   & & {140}  & {0.129}   \\ 
{142} & {0.142}   & & {142}  & {0.192}    \\ 
{144} & {0.203}   & & {144}  & {0.263}  \\ 
{146} & {0.271}   & & {146}  & {0.339}  \\ 
\hline
\hline
\end {tabular}
\end{center}
\label{TabE}
\end{table}

\end{document}